\documentclass[conference]{IEEEtran}
\usepackage{blindtext, graphicx}
\ifCLASSINFOpdf
\else
\fi
\usepackage{algorithm}
\usepackage{algorithmic}

\usepackage[noadjust]{cite}

\begin{document}
%
\title{Millimeter Wave Beam Training: A Survey}

\author{\IEEEauthorblockN{Ish Kumar Jain}
\IEEEauthorblockA{New York University\\
Tandon School of Engineering\\
Email: ishjain@nyu.edu}

}


%


\maketitle

\begin{abstract}
The initial access is required to establish a connection between millimeter wave access point (AP) and users. The large isotropic pathloss at high frequencies can be mitigated by the use of highly directional antennas. However, it complicated the initial access procedures and increased the latency, defying one of the major low latency objective of 5G systems. Beam training algorithms are developed for the AP as well as for the users to find the desired beam quickly and reduce the initial access delay. But, these beam training protocols have to be run very frequently due to outage events, user's mobility and period of sleep cycles. 
\end{abstract}


%
\IEEEpeerreviewmaketitle

\section{Survey}
A multitude of literature is available for initial access techniques at sub-6GHz frequencies in ad-hoc wireless network scenarios or, at 60 GHz IEEE 802.11ad WLAN and WPAN scenarios \cite{wang2009beam,alkhateeb2013hybrid, tsang2011coding,baykas2011ieee,kim2014fast, zhou2012efficient}. However, the initial access for the mmWave cellular network is burgeoning with very recent papers \cite{hur2013millimeter,barati2014directional,desai2014initial,barati2015directional,li2016performance, palacios2016speeding, jeong2015random}. 
The initial access at high frequencies involves a directional search to align the directional beams of the two end users for the maximum gain using a set of procedures known as beam training. It is also referred by different names like initial beamforming, beamforming training or beam steering. We use these terms interchangeably for a complete survey of all of the techniques.

The techniques for beam training can be broadly classified into two categories: (i) Sequential Search and (ii) Hierarchical Search. 
In sequential search, the AP and the UE use a highly directional beam and scan the beam space to find the desired beam. Initial work as in \cite{jeong2015random, barati2014directional} proposed a blind exhaustive search where the AP randomly transmits synchronization signals in different directions for each time slot, eventually scanning the whole angular space. This naive approach might entail a high delay in obtaining the desired beam. 
Thus, multiple proposals have been introduced to reduce the beam training delay. In particular, \cite{zhou2012efficient}, \cite{barati2016initial} and\cite{ barati2015directional} proposed the use of quasi-omnidirectional beam pattern at the transmitter, while the receiver does an exhaustive scan over all possible beam space; this process is reversed in a second phase  with the transmitter scanning the space, while the receiver uses a quasi-omnidirectional beam. 
\cite{li2017design} is a recent paper that compares four different exhaustive search techniques using a combination of directional and omnidirectional transmission.
But, using an omnidirectional antenna in the mmWave cellular environment requires a huge power for a large cellular coverage and causes a lot of interference to other devices.
Another smart approach is the use of multiple directional beams simultaneously. 
For instance, \cite{tsang2011coding} generated multiple beams simultaneously by manipulating the antenna weights and then coded the beams with a unique signature. Agile link beam training \cite{abari2016millimeter} instead used sparse FFT and randomized hashing to generate multiple beams and then used a voting procedure to find the right direction.
However, obtaining multiple beams simultaneously is difficult and it suffers from high power in the side lobes.





On the other hand, the hierarchical search techniques use a combination of high and low-resolution antenna patterns iteratively for beamforming. The AP first performs an exhaustive search on wide beam and then refines to search narrow beams. These protocols follow some codebook for the beam pattern. For instance, the beam coding presented in \cite{wang2009beam} consists of 3 stages---namely, the device to device linking, sector-level searching, and beam-level searching. This is also an optional part of IEEE 802.15.3c license for 60 GHz mmWave WPAN system.
Various other codebook designs are presented in \cite{alkhateeb2013hybrid,hur2013millimeter,kim2014fast,eltayeb2015opportunistic,li2016performance,de2017millimeter}. In particular, \cite{hur2013millimeter} is designed for wireless backhaul between fixed APs and \cite{kim2014fast} is designed for adaptive modulation schemes. Since these techniques use a wide beam during the initial phase, they all suffer from the limited coverage, with the worst performance for the users at the boundary of the cell. Moreover, high power is needed for initial sector-level searching which may result in high interference with other users.


A recent survey \cite{giordani2016comparative} compared the sequential search and hierarchical search techniques for the mmWave cellular network scenario. They showed a trade-off between beam training delay and misdetection probability. The hierarchical search gives a smaller delay, but since they make use of small antenna array in the first phase, they present an order of magnitude higher misdetection probability as compared to the exhaustive search technique. 

Implementing these beam training techniques for mmWave frequencies require some special beamforming architectures.
The fully digital architecture used for sub-6 GHz Massive MIMO systems may provide a small access delay \cite{barati2014directional}, but it is impractical at mmWave frequencies because of the high cost, high power consumption at the analog-to-digital converters, and the complexity of mixed-signal hardware which prevents the use of dedicated RF chain per antenna element. 
On the other hand, analog architecture requires only a single RF chain to process all the antennas. Unlike MIMO systems, it uses a network of phase shifters that controls the phase of each antenna element to produce a directional beam \cite{hur2013millimeter,kim2014fast,li2016performance,abari2016millimeter}. 
A hybrid of digital and analog architecture which requires only a few RF chains is capable of simultaneous multi-direction scanning. Adaptive hierarchical beam training and codebook design for hybrid architecture is considered in \cite{alkhateeb2013hybrid, eltayeb2015opportunistic, palacios2016speeding, de2017millimeter}. In fact, 
\cite{alkhateeb2013hybrid} is the first paper that exploited the sparsity of mmWave channel and the availability of partial channel knowledge and proposed a hybrid iterative beamforming protocol utilizing a variant of matching pursuit algorithm.

A comparison of analog architecture using an exhaustive sequential search procedure with the digital and hybrid architectures for both sequential and hierarchical search procedures is presented in \cite{desai2014initial}. Their simulation reveals that the hierarchical search always performs worse than sequential search. The reason mentioned in the paper is that the wide beams (\textit{e.g.} $45^\circ$) used for hierarchical search had more power in the sidelobes than narrow beams, which resulted in UE selecting an incorrect beam at the initial stage of the hierarchical search procedure. The performance of digital and hybrid architecture was almost similar but worse than the analog architecture with exhaustive search. 
\ifCLASSOPTIONcaptionsoff
  \newpage
\fi



\bibliographystyle{IEEEtran}
%

\bibliography{IEEEbib.bib}




%

\begin{IEEEbiography}[{\includegraphics[width=1in,height=1.25in,clip,keepaspectratio]{picture}}]{John Doe}
\blindtext
\end{IEEEbiography}




\end{document}